\documentclass[aps,preprint]{revtex4}%
\usepackage{amsfonts}
\usepackage{amsmath}
\usepackage{amssymb}
\usepackage{subfigure}
\usepackage{graphicx}%
\begin{document}
\preprint{CTP-SCU/2017037}
\title{Membrane Paradigm and Holographic DC Conductivity for Nonlinear Electrodynamics}
\author{Xiaobo Guo$^{a}$}
\email{guoxiaobo@swust.edu.cn}
\author{Peng Wang$^{b}$}
\email{pengw@scu.edu.cn}
\author{Haitang Yang$^{b}$}
\email{hyanga@scu.edu.cn}
\affiliation{$^{a}$School of Science, Southwest University of Science and Technology,
Mianyang, 621010, PR China }
\affiliation{$^{b}$Center for Theoretical Physics, College of Physical Science and
Technology, Sichuan University, Chengdu, 610064, PR China}

\begin{abstract}
Membrane paradigm is a powerful tool to study properties of black hole
horizons. We first explore the properties of the nonlinear electromagnetic
membrane of black holes. For a general nonlinear electrodynamics field, we
show that the conductivities of the horizon usually have off-diagonal
components and depend on the normal electric and magnetic fields on the
horizon. Via the holographic duality, we find a model-independent expression
for the holographic DC conductivities of the conserved current dual to a probe
nonlinear electrodynamics field in a neutral and static black brane
background. It shows that these DC conductivities only depend on the geometric
and electromagnetic quantities evaluated at the horizon. We can also express
the DC conductivities in terms of the temperature, charge density and magnetic
field in the boundary theory, as well as the values of the couplings in the
nonlinear electrodynamics at the horizon.

\end{abstract}
\maketitle
\tableofcontents

\bigskip



\section{Introduction}

Black holes are among the most intriguing concepts of general relativity. The
event horizon of a black hole is a puzzling and fascinating object, in that
natural descriptions of physics often have trouble accommodating the horizon.
One of the challenges of extending theories to the horizon is defining
boundary conditions on the horizon. The horizon is a null hypersurface, which
possesses a singular Jacobian and a both normal and tangential vector field.
If one believes that $\left(  1\right)  $ the effective number of degrees of
freedom within a few Planck lengths away from the horizon is very small,
$\left(  2\right)  $ the interior of the black hole is a classically
inaccessible region to an outside observer, an effective timelike membrane can
be put just outside the horizon to have boundary conditions defined on it,
instead of the horizon. These observations form the basis of the membrane
paradigm for black holes.

The membrane paradigm was proposed and developed by Thorne and his Caltech
colleagues in a series of papers
\cite{IN-Thorn:1982mn,IN-MacDonald:1982mn,IN-MacDonald:1984ds,IN-Price:1986yy,IN-Suen:1988kq}%
. Later, a more systematic action-based derivation was obtained by Parikh and
Wilczek in \cite{IN-Parikh:1997ma}, which could determine membrane properties
for various field theories. The membrane paradigm was originally developed to
serves as an efficient computational tool useful in dealing with some
astrophysical physics in black hole backgrounds
\cite{IN-Masso:1998fi,IN-Komissarov:2004ms,IN-Penna:2013rga}. On the other
hand, the membrane paradigm is also useful to study microscopic properties of
black hole horizons. For example, the membrane paradigm predicts that black
hole horizons are the fastest scramblers in nature
\cite{IN-Susskind2005sw,IN-Sekino:2008he}. In particular, authors of
\cite{IN-Fischler:2015cma} studied the electromagnetic membrane properties of
the horizon and considered a charged particle dropping onto the horizon in the
framework of Maxwell-Chern-Simons theory. They found that black hole horizon
behaved as a Hall conductor, and there were vortices introduced to the way
perturbations scramble on the horizon.

The AdS/CFT duality \cite{IN-Banks:1996vh,IN-Maldacena:1997re} conjectures a
connection between a strongly coupled gauge theory in $d$-dimensions on the
boundary and a dual weakly coupled gravity in $\left(  d+1\right)  $
dimensional bulk spacetime. Recently, a renewed interest has emerged\ in the
study of membrane paradigm in the context of the AdS/CFT duality. It has been
shown that the membrane paradigm fluid on the black hole horizon is
conjectured to give the low frequency limit of linear response of a strongly
coupled quantum field theory at a finite temperature
\cite{IN-Kovtun:2003wp,IN-Kovtun:2004de,IN-Iqbal:2008by,IN-Bredberg:2010ky}.
In particular, identifying the currents in the boundary theory with radially
independent quantities in bulk, authors of \cite{IN-Iqbal:2008by} showed that
the low frequency limit of the boundary theory transport coefficients could be
expressed in terms of geometric quantities evaluated at the horizon. This
method was later extended to calculate the DC conductivity in the presence of
momentum dissipation
\cite{IN-Blake:2013bqa,IN-Donos:2014cya,IN-Cremonini:2016avj,IN-Bhatnagar:2017twr}%
, where the zero mode of the current, not the current itself, did not evolve
in the radial direction. Specifically, the DC thermoelectric conductivity has
recently been obtained by solving a system of Stokes equations on the black
hole horizon for a charged fluid in Einstein-Maxwell theory
\cite{IN-Donos:2015gia}. Later, the technology of \cite{IN-Donos:2015gia} has
been extended to various theories, e.g. Einstein-Maxwell-scalar theory
\cite{IN-Banks:2015wha} and including a $\theta F\wedge F$ term in the
Lagrangian \cite{IN-Donos:2017mhp}.

Nonlinear electrodynamics (NLED) are effective models incorporating quantum
corrections to Maxwell electromagnetic theory. Among the various NLED, there
are two well-known ones: $\left(  1\right)  $ Heisenberg--Euler effective
Lagrangian that contains logarithmic terms \cite{IN-Heisenberg:1935qt}. These
terms describe the vacuum polarization effects and take into account one-loop
quantum corrections to QED; $\left(  2\right)  $ Born-Infeld electrodynamics
that incorporates maximal electric fields and smooths divergences of the
electrostatic self-energy of point charges \cite{IN-Born:1934gh}. Coupling
NLED to gravity, various NLED charged black holes were derived in a number of
papers
\cite{IN-Soleng:1995kn,IN-AyonBeato:1998ub,IN-Cai:2004eh,IN-Maeda:2008ha,IN-Hendi:2017mgb}%
. Some of these black holes are non--singular exact black hole solutions
\cite{IN-AyonBeato:1998ub}. In the framework of AdS/CFT duality, the shear
viscosity was calculated in Einstein-Born-Infeld gravity \cite{IN-Cai:2008in}.
Holographic superconductors were studied in several NLED-gravity theory
\cite{IN-Jing:2010zp,IN-Jing:2011vz,IN-Gangopadhyay:2012np,IN-Roychowdhury:2012hp}%
. In \cite{IN-Dehyadegari:2017fqo}, the holographic conductivity for the black
brane solutions in the massive gravity with a power-law Maxwell field was
numerically investigated. A class of holographic models for Mott insulators,
whose gravity dual contained NLED, was studied in \cite{IN-Baggioli:2016oju}.
The properties of magnetotransport in holographic Dirac-Born-Infeld models
were discussed in a probe case \cite{IN-Kiritsis:2016cpm} and taking into
account the effects of backreaction on the geometry
\cite{In-Cremonini:2017qwq}.

In this paper, we will consider a neutral and static black brane background
with a probe NLED field and its dual theory. The aim of the paper is to study
the NLED electromagnetic membrane properties and find a model-independent
expression for the holographic DC conductivities of the dual conserved current
in the boundary theory. Specifically, we give a quick review of nonlinear
electromagnetic fields in the curved spacetime in section \ref{Sec:NLED}. In
section \ref{Sec:MP}, we use the membrane paradigm to compute the
conductivities of the stretched horizon. Unlike Maxwell or
Maxwell-Chern-Simons theories, we find that, for a general NLED field, the
conductivities would usually depend on the normal electric and magnetic fields
on the stretched horizon. In section \ref{Sec:ICRS}, we consider a charged
point particle infalling into the horizon in Rindler space. It shows that
effects of NLED do not affect the charge density on the stretched horizon or
the scrambling time, but only could change the way the charge scramble. In
section \ref{Sec:DCFGGD}, the DC conductivities of the dual conserved current
are calculated in the framework of Gauge/Gravity duality. We show that these
DC conductivities usually depend on both the geometry and values of the
couplings in NLED at the black hole horizon as well as the probe charge
density and magnetic field in the boundary theory. In section \ref{Sec:Con},
we conclude with a brief discussion of our results. We use convention that the
Minkowski metric has signature of the metric $\left(  -+++\right)  \,$in this paper,.

\section{Nonlinear Electrodynamics}

\label{Sec:NLED}

In this section, we will briefly review nonlinear electromagnetic fields in
the curved spacetime, mainly in order to define terms and notation and derive
formulas for later use. Let us consider the action of a nonlinear
electromagnetic field $A_{a}$ in a $\left(  3+1\right)  $-dimensional manifold
$\mathcal{M}$%
\begin{equation}
S=\int\limits_{\mathcal{M}}d^{4}x\sqrt{-g}\left[  \mathcal{L}\left(
s,p\right)  +J^{a}A_{a}\right]  , \label{eq:action}%
\end{equation}
where we build two independent nontrivial scalars using $F_{ab}$ and none of
its derivatives:
\begin{align}
s  &  =-\frac{1}{4}F^{ab}F_{ab}\text{,}\nonumber\\
p  &  =-\frac{1}{8}\epsilon^{abcd}F_{ab}F_{cd}\text{;}%
\end{align}
the field strength is defined by $F_{ab}=\partial_{a}A_{b}-\partial_{b}A_{a}$;
$\epsilon^{abcd}\equiv-\left[  a\text{ }b\text{ }c\text{ }d\right]  /\sqrt
{-g}$ is a totally antisymmetric Lorentz tensor, and $\left[  a\text{ }b\text{
}c\text{ }d\right]  $ is the permutation symbol; the Lagrangian density
$\mathcal{L}\left(  s,p\right)  $ is an arbitrary function of $s$ and $p$;
$J^{a}$ is the external current. We also assume that the NLED Lagrangian would
reduce to the form of Maxwell-Chern-Simons Lagrangian for small fields:%
\begin{equation}
\mathcal{L}\left(  s,p\right)  \approx g\left(  x\right)  s+\theta\left(
x\right)  p,
\end{equation}
where, for later convenience, we define%
\begin{equation}
g\left(  x\right)  \equiv\mathcal{L}^{\left(  1,0\right)  }\left(  0,0\right)
\text{ and }\theta\left(  x\right)  \equiv\mathcal{L}^{\left(  0,1\right)
}\left(  0,0\right)  .
\end{equation}
Here we allow coordinate dependent couplings in $\mathcal{L}\left(
s,p\right)  $. The equations of motion obtained from the action $\left(
\ref{eq:action}\right)  $ are%
\begin{equation}
\nabla_{a}G^{ab}+J^{b}=0\text{,} \label{eq:eomS}%
\end{equation}
where we define%
\begin{equation}
G^{ab}=-\frac{\partial\mathcal{L}\left(  s,p\right)  }{\partial F_{ab}}%
=\frac{\partial\mathcal{L}\left(  s,p\right)  }{\partial s}F^{ab}+\frac{1}%
{2}\frac{\partial\mathcal{L}\left(  s,p\right)  }{\partial p}\epsilon
^{abcd}F_{cd}. \label{eq:Gab}%
\end{equation}
Meanwhile, from the definition of $F_{ab}$, the field strength also satisfies
the Bianchi identity%
\begin{equation}
F_{\left[  ab,c\right]  }=\frac{1}{3}\left(  F_{ab,c}+F_{bc,a}+F_{ca,b}%
\right)  =0. \label{eq:eomB}%
\end{equation}

The electric and magnetic fields measured by an observer with $4$-velovity
$U^{a}$ are given by%
\begin{align}
E^{a}  &  =F^{ab}U_{b},\nonumber\\
B^{a}  &  =\frac{1}{2}\epsilon^{bacd}F_{cd}U_{b},
\end{align}
Note that the fields $E^{a}$ and $B^{a}$ are $3$-vectors since $E^{a}%
U_{a}=B^{a}U_{a}=0$. The variables $s$ and $p$ can be rewritten in terms of
$E^{a}$ and $B^{a}$:%
\begin{align}
s  &  =\frac{1}{2}\left(  E^{a}E_{a}-B^{a}B_{a}\right)  ,\nonumber\\
p  &  =-E^{a}B_{a}.
\end{align}
Similarly for $G^{ab}$, we can define auxiliary fields $D^{a}$ and $H^{a}$:%
\begin{align}
D^{a}  &  =G^{ab}U_{b},\nonumber\\
H^{a}  &  =\frac{1}{2}\epsilon^{bacd}G_{cd}U_{b},
\end{align}
which are related to $E^{a}$ and $B^{a}$ by%
\begin{align}
D^{a}  &  =\frac{\partial\mathcal{L}\left(  s,p\right)  }{\partial s}%
E^{a}-\frac{\partial\mathcal{L}\left(  s,p\right)  }{\partial p}%
B^{a},\nonumber\\
H^{a}  &  =\frac{\partial\mathcal{L}\left(  s,p\right)  }{\partial s}%
B^{a}+\frac{\partial\mathcal{L}\left(  s,p\right)  }{\partial p}E^{a}.
\label{eq:EBDH}%
\end{align}
The electromagnetic $4$-current $J^{a}$ can be split into the charge density
$\rho$ and current density $\mathbf{J}^{a}$ measured by the observer:
\begin{align}
\rho &  =-J^{a}U_{a},\nonumber\\
\mathbf{J}^{a}  &  =J^{a}-\sigma U^{a},
\end{align}
where $\mathbf{J}^{a}$ is a 2-vector since $\mathbf{J}^{a}U_{a}=\mathbf{J}%
^{a}n_{a}=0$.

Born-Infeld electrodynamics is described by the Lagrangian density%
\begin{equation}
\mathcal{L}_{\text{BI}}\left(  s,p\right)  =-b^{2}\sqrt{1-\frac{2s}{b^{2}%
}-\frac{p^{2}}{b^{4}}}+b^{2},
\end{equation}
where the coupling parameter $b$ is related to the string tension
$\alpha^{\prime}$ as $b=1/2\pi\alpha^{\prime}$. For small fields $s,p\ll
b^{2}$, we can recover the Maxwell Lagrangian. A simple example of an
electrodynamics Lagrangian with a logarithmic term has the form%
\begin{equation}
\mathcal{L}_{\text{Log}}\left(  s,p\right)  =-b^{2}\log\left(  1-\frac
{s}{b^{2}}\right)  .
\end{equation}
This Lagrangian also reduces to the Maxwell case in the limit $b\rightarrow
\infty$.

As a simple example, let us calculate the electric and magnetic fields of a
point charge in four-dimensional Minkowski space. In Minkowski space, eqns.
$\left(  \ref{eq:eomS}\right)  $ and $\left(  \ref{eq:eomB}\right)  $ become%
\begin{align}
\nabla\times\vec{E}  &  =-\frac{\partial\vec{B}}{\partial t}\text{,}%
\nonumber\\
\nabla\cdot\vec{B}  &  =0\text{,}\nonumber\\
\nabla\cdot\vec{D}  &  =\rho\text{,}\label{eq:eomEB}\\
\nabla\times\vec{H}  &  =\vec{j}+\frac{\partial\vec{D}}{\partial t}%
\text{,}\nonumber
\end{align}
where $\rho=Q\delta^{3}\left(  \vec{r}\right)  $ and $\vec{j}=0$ for a point
charge sitting at $\vec{r}=0$. Since $\partial\vec{E}/\partial t=\partial
\vec{B}/\partial t=0$ in this case, eqns. $\left(  \ref{eq:eomEB}\right)  $
lead to%
\begin{equation}
\vec{D}=\frac{Q}{4\pi r^{2}}\hat{r}\text{ and }B=0\text{.} \label{eq:DB}%
\end{equation}
Considering $s=E^{2}/2$ and $p=0$, we can solve eqns. $\left(  \ref{eq:EBDH}%
\right)  $ for $\vec{E}$ and find that%
\begin{equation}
\vec{E}=\frac{Q}{4\pi r^{2}}F\left(  \frac{Q}{4\pi r^{2}}\right)  \hat{r},
\label{eq:E}%
\end{equation}
where $y\left(  x\right)  =xF\left(  x\right)  $ is the inverse of the
function $x\left(  y\right)  =\mathcal{L}^{\left(  1,0\right)  }\left(
\frac{y^{2}}{2},0\right)  y$. For example, one has%
\begin{equation}
F\left(  x\right)  =\left\{
\begin{array}
[c]{c}%
\frac{1}{\sqrt{1+x^{2}/b^{2}}}\text{ \ \ \ \ \ \ \ Born-Infeld
electrodynamics}\\
\frac{-1+\sqrt{1+2x^{2}/b^{2}}}{x^{2}/b^{2}}\text{ \ \ \ Logarithmic
electrodynamics}%
\end{array}
\right.  .
\end{equation}

\section{Membrane Paradigm}

\label{Sec:MP}

In this section, we will begin with a brief discussion of the action
formulation of the black hole membrane paradigm and then examine the
electromagnetic membrane in the framework of NLED. The interested reader can
find a detailed discussion of the action formulation of the membrane paradigm
in \cite{IN-Parikh:1997ma}.

The black hole has an event horizon, $\mathcal{H}$, which is a 3-dimensional
null hypersurface with a null geodesic generator $l_{a}$. We then choose a
time-like surface just outside $\mathcal{H}$, which is called the stretched
horizon and denoted by $\mathcal{S}$. We regard $\mathcal{S}$ as the
world-tube of a family of fiducial observers, which have world lines $U^{a}$.
The stretched horizon also possesses a spacelike the outward pointing normal
vector $n_{a}$, which satisfies $n^{a}\nabla_{a}n^{c}=0$ on $\mathcal{S}$.
Since the region inside the black hole cannot classically affect an outside
observer $\mathcal{O}$, the classical equations of motion for $\mathcal{O}$
must be obtained by extremizing the action restricted to the spacetime outside
the black hole, $S_{\text{out}}$. However, $S_{\text{out}}$ is not stationary
on its own because there are no boundary conditions fixed at $\mathcal{H}$. To
have the correct Euler-Lagrange equations, it is necessary to add a surface
term $S_{\text{surf}}$ to $S_{\text{out}}$ to exactly cancel all the boundary
terms. In practice, it is often more convenient to define $S_{\text{surf}}$ on
$\mathcal{S}$. Consequently, the total action can be rewritten as
\begin{equation}
S_{\text{tot}}=\left(  S_{\text{out}}+S_{\text{surf}}\right)  +\left(
S_{\text{in}}-S_{\text{surf}}\right)  ,
\end{equation}
where $\delta S_{\text{out}}+\delta S_{\text{surf}}=0$ will give the correct
equations of motion outside $\mathcal{S}$. For the Maxwell action, the surface
term can be interpreted as sources such as surface electric charges and
currents \cite{IN-Parikh:1997ma}.

For a nonlinear electromagnetic field $A_{a}$, the external action is given by
eqn. $\left(  \ref{eq:action}\right)  $:%
\begin{equation}
S_{\text{out}}=\int d^{4}x\sqrt{-g}\left[  \mathcal{L}\left(  s,p\right)
+J^{a}A_{a}\right]  .
\end{equation}
Integration by parts on the kinetic term of the nonlinear electromagnetic
field leads to a variation at the boundary%
\begin{equation}
-\int\limits_{\mathcal{S}}d^{3}x\sqrt{-h}n_{b}G^{ab}\delta A_{a},
\end{equation}
where $h_{ab}=g_{ab}-n_{a}n_{b}$ is the induced metric on $\mathcal{S}$. To
cancel this boundary contribution, we add a surface term $S_{\text{surf}}$%
\begin{equation}
S_{\text{surf}}=\int\limits_{\mathcal{S}}d^{3}x\sqrt{\left\vert h\right\vert
}j_{\text{s}}^{a}A_{a},
\end{equation}
where we define the membrane current as%
\begin{equation}
j_{\text{s}}^{a}=G^{ab}n_{b}. \label{eq:jas}%
\end{equation}
The current $j_{\text{s}}^{a}$ is on the stretched horizon since
$n_{a}j_{\text{s}}^{a}=0$. As in the Maxwell case, this surface term
corresponds to the surface electric charge density $\rho=-j_{\text{s}}%
^{a}U_{a}$ and current density $\mathbf{j}_{\text{s}}^{a}=j_{\text{s}}%
^{a}-\sigma U^{a}$. From eqns. $\left(  \ref{eq:eomS}\right)  $, one can find
a continuity equation for the membrane current $j_{\text{s}}^{a}$:
\begin{equation}
\nabla_{a}j_{\text{s}}^{a}=-J^{a}n_{a},
\end{equation}
where $-J^{a}n_{a}$ describes the charges crossing the stretched horizon.

We now consider a general black brane background, the metric of which takes
the form%
\begin{align}
ds^{2}  &  =g_{ab}dx^{a}dx^{b}=g_{rr}\left(  r\right)  dr^{2}+g_{\mu\nu
}\left(  r\right)  dx^{\mu}dx^{\nu}\nonumber\\
&  =-g_{tt}\left(  r\right)  dt^{2}+g_{rr}\left(  r\right)  dr^{2}%
+g_{AB}\left(  r\right)  dx^{A}dx^{B}\label{eq:metric}\\
&  =-g_{tt}\left(  r\right)  dt^{2}+g_{rr}\left(  r\right)  dr^{2}%
+g_{zz}\left(  r\right)  \left(  dy^{2}+dz^{2}\right)  ,\nonumber
\end{align}
where indices $\left\{  a,b\right\}  $ run over the $\left(  3+1\right)
$-dimensional bulk space, $\left\{  \mu,\nu\right\}  $ over $3$-dimensional
constant-$r$ hypersurface, and $\left\{  A,B\right\}  $ over spatial
coordinates. We assume that there is an event horizon at $r=r_{h}$, where
$g_{tt}\left(  r\right)  $ has a first order zero, $g_{rr}\left(  r\right)  $
has a first order pole, and $g_{zz}\left(  r\right)  $ is nonzero and finite.
The Hawking temperature of the metric $\left(  \ref{eq:metric}\right)  $ is
\begin{equation}
T=\frac{\sqrt{g_{tt}^{\prime}\left(  r_{h}\right)  g^{rr\prime}\left(
r_{h}\right)  }}{4\pi}. \label{eq:HT}%
\end{equation}
We also assume that the couplings in $\mathcal{L}\left(  s,p\right)  $ only
depend on $r$.

Now put the stretched horizon at $r=r_{0}$ with $r_{0}-r_{h}\ll r_{h}$. This
stretched horizon would have%
\begin{equation}
n_{a}=\sqrt{g_{rr}\left(  r_{0}\right)  }\delta_{ar}\text{ and }U_{a}%
=-\sqrt{g_{tt}\left(  r_{0}\right)  }\delta_{at}.
\end{equation}
Thus, the membrane current $\left(  \ref{eq:jas}\right)  $ reduces to%
\begin{equation}
j_{\text{s}}^{\mu}=\sqrt{g_{rr}\left(  r_{0}\right)  }G^{\mu r} \label{eq:mur}%
\end{equation}
To find relations among $F_{ab}\left(  r_{0}\right)  $, we consider a radially
freely falling observer in our background. It is easy to obtain the
$4$-velocity vector of this infalling observer:
\begin{equation}
U_{\text{FFO}}^{a}=\left(  \tilde{E}g_{tt}^{-1},-\tilde{E}g_{tt}^{-1}%
\sqrt{\frac{g_{tt}}{g_{rr}}}\sqrt{1-\tilde{E}^{-2}g_{tt}},0,0\right)  ,
\end{equation}
where $\tilde{E}$ is the conserved energy per unit mass. The fact that $\tau$
is the proper time along the infalling world lines means that $U_{\text{FFO}%
,a}$ is equal to the gradient of $\tau$%
\begin{equation}
U_{\text{FFO},a}=-\partial_{\mu}\tau,
\end{equation}
from which one finds%
\begin{equation}
\frac{\partial\tau}{\partial t}=\tilde{E}\text{ and }\frac{\partial\tau
}{\partial r}=\tilde{E}\sqrt{\frac{g_{rr}}{g_{tt}}}\sqrt{1-\tilde{E}%
^{-2}g_{tt}}.
\end{equation}
Since this freely falling observer does not see the coordinate singularity at
the horizon, the field strength observed by this observer must be regular.
Relating $F^{\tau A}$\ to $F^{rA}$\ and $F^{tA}$, we obtain%
\begin{equation}
F^{\tau A}=\frac{\partial\tau}{\partial r}F^{rA}+\frac{\partial\tau}{\partial
t}F^{tA}\Longrightarrow\tilde{E}\left(  \sqrt{g_{rr}}\sqrt{1-\tilde{E}%
^{-2}g_{tt}}F^{rA}+\sqrt{g_{tt}}F^{tA}\right)  =F^{\tau A}\sqrt{g_{tt}},
\label{eq:taoA}%
\end{equation}
where $F^{\tau A}$ is finite at $r=r_{h}$. On the stretched horizon,
$\sqrt{g_{tt}}F^{tA}$ is the electric field measured by the fiducial
observers, hence it would not go to zero as $r_{0}\rightarrow r_{h}$. Since
$F^{\tau A}\sqrt{g_{tt}}\,$goes to zero as $r_{0}\rightarrow r_{h}$, eqn.
$\left(  \ref{eq:taoA}\right)  $ leads to%
\begin{equation}
F^{rA}\left(  r_{0}\right)  =-\sqrt{\frac{g_{tt}\left(  r_{0}\right)  }%
{g_{rr}\left(  r_{0}\right)  }}F^{tA}\left(  r_{0}\right)  \text{,}
\label{eq:rt}%
\end{equation}
for $r_{0}-r_{h}\ll r_{h}$.

Using eqns. $\left(  \ref{eq:Gab}\right)  $, $\left(  \ref{eq:mur}\right)  $
and $\left(  \ref{eq:rt}\right)  $, we find that%
\begin{equation}
j_{\text{s}}^{A}=\mathcal{L}^{\left(  1,0\right)  }\left(  s,p\right)
|_{r=r_{0}}E^{A}-\left[  A\text{ }B\right]  \mathcal{L}^{\left(  0,1\right)
}\left(  s,p\right)  |_{r=r_{0}}E^{B}, \label{eq:mcurrent}%
\end{equation}
where $\mathcal{L}^{\left(  1,0\right)  }\left(  s,p\right)  =\partial
\mathcal{L}\left(  s,p\right)  /\partial s$ and $\mathcal{L}^{\left(
0,1\right)  }\left(  s,p\right)  =\partial\mathcal{L}\left(  s,p\right)
/\partial p$; the electric and magnetic fields measured by the fiducial
observers on the stretched horizon are%
\[
E^{a}=F^{ta}\left(  r_{0}\right)  \sqrt{g_{tt}\left(  r_{0}\right)  }\text{
and }B^{a}\left(  r_{0}\right)  =\frac{1}{2}\frac{\left[  t\text{ }a\text{
}c\text{ }d\right]  }{\sqrt{g_{rr}\left(  r_{0}\right)  }g_{zz}\left(
r_{0}\right)  }F_{cd}\left(  r_{0}\right)  \text{, respectively;}%
\]
two variables $s$ and $p$ on the stretched horizon become%
\begin{equation}
s\left(  r_{0}\right)  =\frac{1}{2}\left(  E^{r}E_{r}-B^{r}B_{r}\right)
\text{ and }p\left(  r_{0}\right)  =-E^{r}B_{r}.
\end{equation}
Since $s$ and $p$ are scalars and the field strength observed by the freely
falling observer is regular on the horizon, $s\left(  r_{0}\right)  $ and
$p\left(  r_{0}\right)  $ stay finite as $r_{0}\rightarrow r_{h}$.

The conductivities of the stretched horizon can be read off from eqn. $\left(
\ref{eq:mcurrent}\right)  $:%
\begin{equation}
\sigma_{\text{s}}^{yy}=\sigma_{\text{s}}^{zz}=\mathcal{L}^{\left(  1,0\right)
}\left(  s,p\right)  |_{r=r_{0}}\text{ and }\sigma_{\text{s}}^{zy}%
=-\sigma_{\text{s}}^{yz}=\mathcal{L}^{\left(  0,1\right)  }\left(  s,p\right)
|_{r=r_{0}},
\end{equation}
where $\sigma^{zy}$ is the surface Hall conductance. In Maxwell-Chern-Simons
theory with $\mathcal{L}\left(  s,p\right)  =s+\theta p$, one has
\begin{equation}
\sigma_{\text{s}}^{yy}=\sigma_{\text{s}}^{zz}=1\text{ and }\sigma_{\text{s}%
}^{zy}=-\sigma_{\text{s}}^{yz}=\theta,
\end{equation}
which agree with what was found in \cite{IN-Fischler:2015cma}. However in NLED
models, the conductivities of the stretched horizon usually depend on the
external fields through $s\left(  r_{0}\right)  $ and $p\left(  r_{0}\right)
$. In particular, the conductivities only depend on the normal electric and
magnetic fields measured by the fiducial observers on the stretched horizon.
Note that the normal components of the electric and magnetic fields in an
orthonormal frame of fiducial observers are the same as in those of freely
falling observers. It is noteworthy that the surface charge density
$\rho_{\text{s}}$ of the stretched horizon can be related to $E_{r}$ and
$B_{r}$ via%
\begin{equation}
\rho_{\text{s}}=n_{a}D^{a}=\sqrt{g_{rr}\left(  r_{0}\right)  }\left[
\mathcal{L}^{\left(  1,0\right)  }\left(  s,p\right)  |_{r=r_{0}}%
E_{r}-\mathcal{L}^{\left(  0,1\right)  }\left(  s,p\right)  |_{r=r_{0}}%
B_{r}\right]  .
\end{equation}
Using this equation, we can rewrite the conductivities in terms of the surface
charge density and normal magnetic field.

\section{Infalling Charge in Rindler Space}

\label{Sec:ICRS}

In this section, we will consider dropping a charged point particle into the
horizon in Rindler space. Rindler space is a good approximation to the
Schwarzschild geometry in the near-horizon region $r-2M\ll2M$ and ignores the
spatial curvature there. The metric of Rindler space takes form%
\begin{equation}
ds^{2}=-r^{2}d\omega^{2}+dr^{2}+dy^{2}+dz^{2},
\end{equation}
which describes a portion of Minkowski space, namely the Rindler wedge.
Minkowski coordinates $t$ and $x$ can be defined by%
\begin{equation}
t=r\sinh\omega\text{ and }x=r\cosh\omega
\end{equation}
to get the familiar Minkowski metric%
\begin{equation}
ds^{2}=-dt^{2}+dx^{2}+dy^{2}+dz^{2}\text{.}%
\end{equation}
The Rindler coordinate has a coordinate singularity at $r=0$, where the
determinant of the metric tensor becomes zero. In fact, there is an event
horizon at $r=0$, which becomes $x=\left\vert t\right\vert $ in Minkowski
coordinates and is the edge of the Rindler wedge. We will put the stretched
horizon at $r=r_{0}\ll1$, which has%
\begin{equation}
n^{a}=\left(  0,1,0,0\right)  \text{ and }U^{a}=\left(  r_{0}^{-1}%
,0,0,0\right)  .
\end{equation}

Without loss of generality, we can take a single charge to be at static at
position $x=a$ in Minkowski coordinates. In the Rindler coordinates, the
charge is freely falling into the horizon. In this case, the magnetic and
electric fields in Minkowski coordinates have been obtained in section
\ref{Sec:NLED} and are given by eqns. $\left(  \ref{eq:DB}\right)  $ and
$\left(  \ref{eq:E}\right)  $ with $r^{2}=\left(  x-a\right)  ^{2}+y^{2}%
+z^{2}$, respectively. In Minkowski coordinates, the field strength is%
\begin{align}
F_{\text{M}}^{tx}  &  =-F_{\text{M}}^{xt}=\frac{Q\left(  x-a\right)  }%
{4\pi\left[  \left(  x-a\right)  ^{2}+y^{2}+z^{2}\right]  ^{\frac{3}{2}}%
}F\left(  \frac{Q}{4\pi\left[  \left(  x-a\right)  ^{2}+y^{2}+z^{2}\right]
}\right)  ,\nonumber\\
F_{\text{M}}^{ty}  &  =-F_{\text{M}}^{yt}=\frac{Qy}{4\pi\left[  \left(
x-a\right)  ^{2}+y^{2}+z^{2}\right]  ^{\frac{3}{2}}}F\left(  \frac{Q}%
{4\pi\left[  \left(  x-a\right)  ^{2}+y^{2}+z^{2}\right]  }\right)  ,\\
F_{\text{M}}^{tz}  &  =-F_{\text{M}}^{zt}=\frac{Qz}{4\pi\left[  \left(
x-a\right)  ^{2}+y^{2}+z^{2}\right]  ^{\frac{3}{2}}}F\left(  \frac{Q}%
{4\pi\left[  \left(  x-a\right)  ^{2}+y^{2}+z^{2}\right]  }\right)  ,\nonumber
\end{align}
and all the other components are zero. Performing the change of coordinates to
calculate $F_{\text{R}}^{ab}$ leads to
\begin{align}
j_{\text{s}}^{\omega}  &  =\frac{Q\left(  r_{0}\cosh\omega-a\right)  }{4\pi
r_{0}\left[  \left(  r_{0}\cosh\omega-a\right)  ^{2}+y^{2}+z^{2}\right]
^{\frac{3}{2}}},\nonumber\\
j_{\text{s}}^{\rho}  &  =0,\nonumber\\
j_{\text{s}}^{y}  &  =\frac{Q}{4\pi\left[  \left(  r_{0}\cosh\omega-a\right)
^{2}+y^{2}+z^{2}\right]  ^{\frac{3}{2}}}\left[  y\sinh\omega-z\cosh\omega
\frac{\mathcal{L}^{\left(  0,1\right)  }\left(  s,0\right)  }{\mathcal{L}%
^{\left(  1,0\right)  }\left(  s,0\right)  }|_{r=r_{0}}\right]  ,\\
j_{\text{s}}^{z}  &  =\frac{Q}{4\pi\left[  \left(  r_{0}\cosh\omega-a\right)
^{2}+y^{2}+z^{2}\right]  ^{\frac{3}{2}}}\left[  z\sinh\omega+y\cosh\omega
\frac{\mathcal{L}^{\left(  0,1\right)  }\left(  s,0\right)  }{\mathcal{L}%
^{\left(  1,0\right)  }\left(  s,0\right)  }|_{r=r_{0}}\right] \nonumber
\end{align}
where we use $F\left(  X\right)  \mathcal{L}^{\left(  1,0\right)  }\left(
\frac{X^{2}}{2}F^{2}\left(  X\right)  ,0\right)  =1$ to eliminate the function
$F\left(  X\right)  $, and
\begin{equation}
s\left(  r_{0}\right)  =\frac{Q^{2}}{32\pi^{2}\left[  \left(  r_{0}\cosh
\omega-a\right)  ^{2}+y^{2}+z^{2}\right]  ^{2}}F^{2}\left(  \frac{Q}%
{4\pi\left[  \left(  r_{0}\cosh\omega-a\right)  ^{2}+y^{2}+z^{2}\right]
}\right)  .
\end{equation}
For Maxwell-Chern-Simons theory, we also reproduce the results in
\cite{IN-Fischler:2015cma}. An fiducial observer will measure a surface charge
density $\rho_{\text{s}}=-j_{\text{s}}^{a}U_{a}=r_{0}j_{\text{s}}^{\omega}$.
In NLED, the surface charge density $\rho_{\text{s}}$ of the stretched horizon
is exactly the same as in Maxwell theory. In NLED, there is no corrections to
$\rho_{\text{s}}$, but the surface currents $j_{\text{s}}^{y}$ and
$j_{\text{s}}^{z}$ could receive corrections.

Let us study scrambling of a point charges on the stretched horizon for large
Rindler time. When $\omega\gg1$, we obtain
\begin{align*}
\rho_{\text{s}}  &  =\frac{r_{0}Qe^{-2\omega}}{\pi\left(  r_{0}^{2}+r_{\perp
}^{2}\right)  ^{\frac{3}{2}}},\\
j_{\text{s}}^{y}  &  =\frac{Q}{\pi\left(  r_{0}^{2}+r_{\perp}^{2}\right)
^{\frac{3}{2}}}\left[  y-\frac{\theta\left(  r_{0}\right)  }{g\left(
r_{0}\right)  }z\right] \\
j_{\text{s}}^{z}  &  =\frac{Q}{\pi\left(  r_{0}^{2}+r_{\perp}^{2}\right)
^{\frac{3}{2}}}\left[  z+\frac{\theta\left(  r_{0}\right)  }{g\left(
r_{0}\right)  }y\right]
\end{align*}
where $r_{\perp}^{2}=4e^{-2\omega}\left(  y^{2}+z^{2}\right)  $, and we use
$s\left(  r_{0}\right)  \sim e^{-2\omega}$ in this limit. Whenever
$\theta\left(  r_{0}\right)  \equiv\mathcal{L}^{\left(  0,1\right)  }\left(
0,0\right)  |_{r=r_{0}}\neq0$, effects of NLED would change the way the charge
scramble but not the scrambling time. In this case, there is the presence of
vortices on the stretched horizon \cite{IN-Fischler:2015cma}.

\section{DC Conductivity From Gauge/Gravity Duality}

\label{Sec:DCFGGD}

Under the long wavelength and low frequency limit, it is expected that there
are connections between the near-horizon region geometry of the bulk gravity
and the dual field theory living on the boundary. Observing that the currents
in the boundary theory could be identified with radially independent
quantities in the bulk, authors of \cite{IN-Iqbal:2008by} found that the low
frequency limit of linear response of the boundary theory could be determined
by the membrane paradigm fluid on the stretched horizon. In particular, they
derived expression for the DC conductivity of the boundary theory in terms of
geometric quantities evaluated at the horizon. In \cite{IN-Iqbal:2008by}, the
conserved current in the boundary theory was dual to a Maxwell field in the
bulk. In this section, we will follow the method in \cite{IN-Iqbal:2008by} to
calculate the DC conductivities of the conserved current in the boundary
theory, which is dual to a NLED field in bulk.

We now consider a probe NLED field in the background of a $\left(  3+1\right)
$-dimensional black brane with the metric $\left(  \ref{eq:metric}\right)  $.
For simplicity, we assume that this black brane is electrically neutral with
trivial background configuration of the NLED field. This black brane describes
an equilibrium state at finite temperature $T$, which is given by eqn.
$\left(  \ref{eq:HT}\right)  $. The action of the NLED field is
\begin{equation}
S=\int d^{4}x\sqrt{-g}\mathcal{L}\left(  s,p\right)  . \label{eq:actionN}%
\end{equation}
This NLED field is a U$\left(  1\right)  $ gauge field and dual to a conserved
current $\mathcal{J}^{\mu}$ in the boundary theory. The corresponding AC
conductivities are given by%
\begin{equation}
\left\langle \mathcal{J}^{A}\left(  k_{\mu}\right)  \right\rangle =\sigma
^{AB}\left(  k_{\mu}\right)  F_{Bt}\left(  r_{B}\right)  , \label{eq:sigma}%
\end{equation}
where the boundary theory lives at $r=r_{B}\rightarrow\infty$. The DC
conductivity is obtained in the long wavelength and low frequency limit:
\begin{equation}
\sigma^{AB}=\lim_{\omega\rightarrow0}\lim_{\vec{k}\rightarrow0}\sigma
^{AB}\left(  k_{\mu}\right)  .
\end{equation}

Apart from varying the action, we can also derive the equations of motion
using Hamiltonian formulation. Using gauge choice $A_{r}=0$, we will find the
equations of motion for $A_{\mu}$ in a Hamiltonian form. From the action
$\left(  \ref{eq:actionN}\right)  $, the conjugate momentum of the field
$A_{\mu}$ with respect to $r$-foliation is given by%
\begin{equation}
\Pi^{\mu}=\frac{\delta S}{\delta\left(  \partial_{r}A_{\mu}\right)  }%
=-\sqrt{-g}G^{r\mu}, \label{eq:pi}%
\end{equation}
where $G^{r\mu}$ are defined in eqn. $\left(  \ref{eq:Gab}\right)  $. Since
$G^{r\mu}$ are functions of $F^{r\mu}$ and $F^{\mu\nu}$, one could solve eqn.
$\left(  \ref{eq:pi}\right)  $ to find an expression for $F^{r\mu}$ in terms
of $\Pi^{\mu}$ and $F^{\mu\nu}$:%
\begin{equation}
F^{r\mu}=F^{r\mu}\left(  \Pi^{\nu},F^{\rho\sigma}\right)  \text{,}
\label{eq:Frmu}%
\end{equation}
where, as will be shown later, the exact form of the function $F^{r\mu}\left(
\Pi^{\nu},F^{\rho\sigma}\right)  $ is not important for our discussion. So the
Hamiltonian is given by%
\begin{equation}
H=\int d^{4}x\sqrt{-g}\mathcal{H}\left(  \Pi^{\nu},F^{\rho\sigma}\right)
=\int d^{4}x\sqrt{-g}\left[  F_{r\mu}\left(  \Pi^{\nu},F^{\rho\sigma}\right)
\Pi^{\mu}-\mathcal{L}\left(  s,p\right)  \right]  ,
\end{equation}
where we use eqn. $\left(  \ref{eq:Frmu}\right)  $ to rewrite $s$ and $p$ in
terms of $\Pi^{\nu}$ and $F^{\rho\sigma}$. Varying the Hamiltonian with
respect to $A_{\mu}$, we write the equations of motion for $A_{\mu}$ in a
Hamiltonian form as%
\begin{equation}
\partial_{r}\Pi^{\mu}=-2\sqrt{-g}\partial_{\nu}\left[  \frac{\partial
\mathcal{H}\left(  \Pi^{\eta},F^{\rho\sigma}\right)  }{\partial F_{\nu\mu}%
}\right]  . \label{eq:pir}%
\end{equation}
Moreover, the Bianchi identity gives%
\begin{equation}
\partial_{r}F_{\mu\nu}+\partial_{\mu}F_{\nu r}\left(  \Pi^{\eta},F^{\rho
\sigma}\right)  +\partial_{\nu}F_{r\mu}\left(  \Pi^{\eta},F^{\rho\sigma
}\right)  =0\text{.} \label{eq:Fmunvr}%
\end{equation}
In the long wavelength and low frequency limit, i.e. $\omega\rightarrow0$ and
$\vec{k}\rightarrow0$ with $F^{\rho\sigma}$ and $\Pi^{\eta}$ fixed, eqns.
$\left(  \ref{eq:pir}\right)  $ and $\left(  \ref{eq:Fmunvr}\right)  $ become%
\begin{equation}
\partial_{r}\Pi^{\mu}=0\text{ and }\partial_{r}F_{\mu\nu}=0.
\end{equation}

Now we discuss boundary conditions for $F_{ab}$ on the stretched horizon at
$r=r_{0}\rightarrow r_{h}$ and the boundary of bulk at $r=r_{B}\rightarrow
\infty$. On the stretched horizon, $s$ and $p$ become%
\begin{align}
s\left(  r_{0}\right)   &  =\frac{1}{2}\left[  g_{rr}\left(  r_{0}\right)
g_{tt}\left(  r_{0}\right)  F^{rt}\left(  r_{0}\right)  ^{2}-\frac{F_{yz}^{2}%
}{g_{zz}^{2}\left(  r_{0}\right)  }\right]  ,\nonumber\\
p\left(  r_{0}\right)   &  =\frac{g_{rr}\left(  r_{0}\right)  g_{tt}\left(
r_{0}\right)  }{\sqrt{-g\left(  r_{0}\right)  }}F^{rt}\left(  r_{0}\right)
F_{yz}, \label{eq:sp}%
\end{align}
where we use eqn. $\left(  \ref{eq:rt}\right)  $ to express $F^{rA}\left(
r_{0}\right)  $ in terms of $F^{tA}\left(  r_{0}\right)  $, and $F_{yz}$ is an
$r$-independent quantity. Using eqns. $\left(  \ref{eq:mur}\right)  $ and
$\left(  \ref{eq:pi}\right)  $, one can relate $\Pi^{A}\left(  r_{0}\right)  $
to $j_{\text{s}}^{A}$:
\begin{equation}
\Pi^{A}\left(  r_{0}\right)  =\sqrt{g_{tt}\left(  r_{0}\right)  }g_{zz}\left(
r_{0}\right)  j_{s}^{A}=\mathcal{L}^{\left(  1,0\right)  }\left(  s,p\right)
|_{r=r_{0}}F_{At}-\left[  A\text{ }B\right]  \mathcal{L}^{\left(  0,1\right)
}\left(  s,p\right)  |_{r=r_{0}}F_{Bt}, \label{eq:piI0}%
\end{equation}
where $F_{At}$ is also $r$-independent. On the boundary of bulk, it showed in
\cite{IN-Iqbal:2008by,DCFGGD-Hartnoll:2009sz} that one point function of
$\mathcal{J}^{A}$ in the presence of source $F_{\mu v}$ can be written as%
\begin{equation}
\left\langle \mathcal{J}^{\mu}\right\rangle =\Pi^{\mu}\left(  r_{B}\right)  .
\label{eq:Jpi}%
\end{equation}
Since $\Pi^{A}$ and $F_{At}$ are $r$-independent in zero momentum limit, we
can use eqns. $\left(  \ref{eq:piI0}\right)  $ and $\left(  \ref{eq:Jpi}%
\right)  $ to show that
\begin{equation}
\left\langle \mathcal{J}^{A}\left(  k_{\mu}\rightarrow0\right)  \right\rangle
=\mathcal{L}^{\left(  1,0\right)  }\left(  s,p\right)  |_{r=r_{0}}%
F_{At}\left(  k_{\mu}\rightarrow0\right)  -\left[  A\text{ }B\right]
\mathcal{L}^{\left(  0,1\right)  }\left(  s,p\right)  |_{r=r_{0}}F_{Bt}\left(
k_{\mu}\rightarrow0\right)  . \label{eq;J}%
\end{equation}
Comparing eqn. $\left(  \ref{eq:sigma}\right)  $ with eqn. $\left(
\ref{eq;J}\right)  $, we can read off the DC conductivities in the dual
theory:%
\begin{equation}
\sigma^{yy}=\sigma^{zz}=\mathcal{L}^{\left(  1,0\right)  }\left(  s,p\right)
|_{r=r_{h}}\text{ and }\sigma^{zy}=-\sigma^{yz}=\mathcal{L}^{\left(
0,1\right)  }\left(  s,p\right)  |_{r=r_{h}}, \label{eq:Conrh}%
\end{equation}
where we take the limit $r\rightarrow r_{h}$. Note that the DC conductivity in
NLED just with $s$ was also obtained in \cite{IN-Baggioli:2016oju}, where
their eqn. $\left(  31\right)  $ in the probe case agrees with our expression
for $\sigma^{yy}$ in eqns. $\left(  \ref{eq:Conrh}\right)  $.

To express $F^{rt}\left(  r_{h}\right)  $ in terms of quantities in the
boundary theory, we can use the following formula%
\begin{equation}
\Pi^{t}\left(  r_{h}\right)  =\Pi^{t}\left(  r\rightarrow\infty\right)
=\left\langle \mathcal{J}^{0}\right\rangle =\rho, \label{eq:rho}%
\end{equation}
where $\rho$ can be interpreted as the charge density in the dual field
theory. Eqn. $\left(  \ref{eq:rho}\right)  $ becomes%
\begin{equation}
\sqrt{\eta}g_{zz}\left(  r_{h}\right)  \mathcal{L}^{\left(  1,0\right)
}\left(  s,p\right)  |_{r=r_{h}}F^{rt}\left(  r_{h}\right)  +\mathcal{L}%
^{\left(  0,1\right)  }\left(  s,p\right)  |_{r=r_{h}}F_{yz}=-\rho,
\label{eq:rhoF}%
\end{equation}
where eqns. $\left(  \ref{eq:sp}\right)  $ give%
\begin{align}
s\left(  r_{h}\right)   &  =\frac{1}{2}\left[  \eta F^{rt}\left(
r_{h}\right)  ^{2}-\frac{F_{yz}^{2}}{g_{zz}^{2}\left(  r_{h}\right)  }\right]
,\nonumber\\
p\left(  r_{h}\right)   &  =\frac{\sqrt{\eta}}{g_{zz}\left(  r_{h}\right)
}F^{rt}\left(  r_{h}\right)  F_{yz}.
\end{align}
One could solve eqn. $\left(  \ref{eq:rhoF}\right)  $ to express
$F^{rt}\left(  r_{h}\right)  $ in terms of $\rho$ and $F_{yz}$ and the plug
this expression into eqn. $\left(  \ref{eq:Conrh}\right)  $ to write
$\sigma^{AB}$ in terms of $\rho$ and $F_{yz}$. Note $F_{yz}$ can be treated as
the magnetic field in the $\left(  2+1\right)  $ dimensional boundary theory,
in which the magnetic field is a scalar filed. .

\begin{figure}[tb]
\begin{center}
\subfigure[{Plot of $\sigma^{yy}$ against $\rho$ and $F_{yz}$ for Born-Infeld electrodynamics}]{
\includegraphics[width=0.32\textwidth]{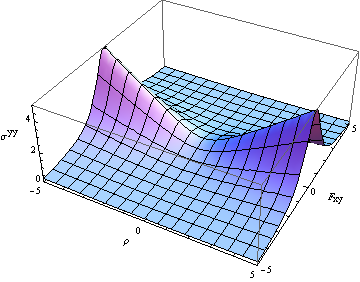}\label{fig:mr:a}}
\subfigure[{Plot of $\sigma^{yz}$ against $\rho$ and $F_{yz}$ for Born-Infeld electrodynamics}]{
\includegraphics[width=0.32\textwidth]{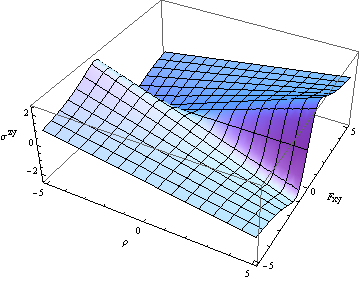}\label{fig:mr:b}}
\subfigure[{Plot of $\sigma^{yy}$ against $\rho$ and $F_{yz}$ for Logarithmic electrodynamics}]{
\includegraphics[width=0.32\textwidth]{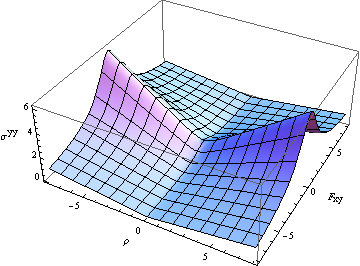}\label{fig:mr:c}}
\end{center}
\caption{Plots of the DC conductivities of the conserved current dual to the
bulk electromagnetic field in Born-Infeld and\ Logarithmic electrodynamics.
Here we set $b^{2}g_{zz}^{2}\left(  r_{h}\right)  =1$.}%
\label{fig:mr}%
\end{figure}

Let us consider $\sigma^{AB}$ in Born-Infeld and\ Logarithmic electrodynamics
discussed above. We find that for Born-Infeld electrodynamics,
\[
\sigma^{yy}=\sigma^{zz}=\frac{\sqrt{1+F_{yz}^{2}/\left[  b^{2}g_{zz}%
^{2}\left(  r_{h}\right)  \right]  +\rho^{2}/\left[  b^{2}g_{zz}^{2}\left(
r_{h}\right)  \right]  }}{1+F_{yz}^{2}/\left[  b^{2}g_{zz}^{2}\left(
r_{h}\right)  \right]  },
\]%
\begin{equation}
\sigma^{yz}=-\sigma^{zy}=\frac{\rho F_{yz}/\left[  b^{2}g_{zz}^{2}\left(
r_{h}\right)  \right]  }{1+F_{yz}^{2}/\left[  b^{2}g_{zz}^{2}\left(
r_{h}\right)  \right]  }, \label{eq:sigmaB}%
\end{equation}
and for Logarithmic electrodynamics%
\[
\sigma^{yy}=\sigma^{zz}=\frac{1+\sqrt{1+\left\{  2+F_{yz}^{2}/\left[
b^{2}g_{zz}^{2}\left(  r_{h}\right)  \right]  \right\}  \rho^{2}/\left[
g_{zz}^{2}\left(  r_{h}\right)  b^{2}\right]  }}{2+F_{yz}^{2}/\left[
b^{2}g_{zz}^{2}\left(  r_{h}\right)  \right]  },
\]%
\begin{equation}
\sigma^{zy}=-\sigma^{yz}=0.
\end{equation}
Note that the DC conductivity matrix of a holographic Dirac-Born-Infeld model
in the probe limit has been calculated in \cite{IN-Kiritsis:2016cpm}. The
eqns. $\left(  3.1\right)  $ and $\left(  3.2\right)  $ with $S=0$ in
\cite{IN-Kiritsis:2016cpm} turn out to be the same as our results $\left(
\ref{eq:sigmaB}\right)  $. In FIG. \ref{fig:mr}, we plot the DC conductivities
versus $\rho$ and $F_{yz}$, of the conserved current dual to the bulk
electromagnetic field in both Born-Infeld and\ Logarithmic
electrodynamics.\ The parameter $b^{2}g_{zz}^{2}\left(  r_{h}\right)  $ sets a
scale in the dual field theory. When $\rho^{2},F_{yz}^{2}\ll b^{2}g_{zz}%
^{2}\left(  r_{h}\right)  $, we practically reproduce the results for Maxwell
theory. On the other hand, effects of nonlinearity of the electromagnetic
fields start to play an important role when $\rho^{2}$ or $F_{yz}^{2}$ are
around the scale $b^{2}g_{zz}^{2}\left(  r_{h}\right)  $. At zero charge
density, the diagonal components of the DC conductivities in both Born-Infeld
and\ Logarithmic electrodynamics are non-zero. These non-zero values can be
interpreted as incoherent contributions \cite{DCFGGD-Davison:2015bea}, known
as the charge conjugation symmetric terms, which are independent of the charge
density $\rho$. As shown in FIG. \ref{fig:mr}, the diagonal DC conductivity
$\sigma^{yy}$ increases with increasing $\left\vert \rho\right\vert $ at
constant $F_{yz}$, which is a feature similar to the Drude metal. For the
Drude metal, a larger charge density provides more available mobile charge
carriers to efficiently transport charge. At constant $\rho$, $\sigma^{yy}$
decreases with increasing $\left\vert F_{yz}\right\vert $, which means a
positive magneto-resistance. 

Since $r_{h}$ is related to the Hawking temperature $T$ by eqn. $\left(
\ref{eq:HT}\right)  $, we can now discuss the temperature dependence of the
conductivities. For simplicity and concreteness, we consider the Schwarzschild
AdS black brane%
\begin{equation}
ds^{2}=-\left(  r^{2}-r_{h}^{3}/r\right)  dt^{2}+\frac{dr^{2}}{\left(
r^{2}-r_{h}^{3}/r\right)  }+r^{2}\left(  dy^{2}+dz^{2}\right)  ,
\end{equation}
where we take the AdS radius $L=1$, and $r_{h}$ determines the Hawking
temperature of the black brane:%
\begin{equation}
T=\frac{3r_{h}}{4\pi}\text{.}%
\end{equation}
Therefore, we obtain that for Born-Infeld electrodynamics,
\begin{align}
\sigma^{yy}  &  =\sigma^{zz}=\frac{\sqrt{1+\lambda F_{yz}^{2}/T^{4}%
+\lambda\rho^{2}/T^{4}}}{\lambda F_{yz}^{2}/T^{4}+1}\\
\sigma^{yz}  &  =-\sigma^{zy}=\frac{\lambda\rho F_{yz}/T^{4}}{1+\lambda
F_{yz}^{2}/T^{4}},
\end{align}
and for Logarithmic electrodynamics%
\begin{equation}
\sigma^{yy}=\sigma^{zz}=\frac{1+\sqrt{1+\left(  2+\lambda F_{yz}^{2}%
/T^{4}\right)  \lambda\rho^{2}/T^{4}}}{2+\lambda F_{yz}^{2}/T^{4}},
\end{equation}
where $\lambda\equiv b^{-2}\left(  4\pi/3\right)  ^{-4}$ is a parameter
associated with the conserved current $\mathcal{J}^{\mu}$ in the boundary
theory. In the high temperature limit, we would recover the results for
Maxwell theory. When $T^{4}\ll\rho^{2}/b^{2}$ and $F_{yz}^{2}/b^{2}$, the low
temperature behavior of $\sigma^{AB}$ is%
\begin{equation}
\sigma^{yy}=\sigma^{zz}\sim T^{2}\text{ and }\sigma^{zy}=-\sigma^{yz}\sim
T^{0}\text{ for Born-Infeld electrodynamics,}%
\end{equation}
and%
\begin{equation}
\sigma^{yy}=\sigma^{zz}\sim T^{0}\text{ for Logarithmic electrodynamics.}%
\end{equation}

\begin{figure}[tb]
\begin{center}
\includegraphics[width=0.5\textwidth]{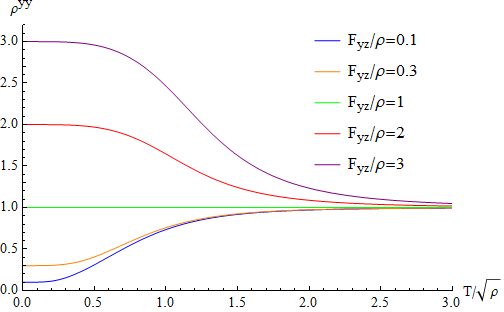}
\end{center}
\caption{Plot of $\rho^{yy}$ versus $T/\sqrt{\rho}$ in Logarithmic
electrodynamics.}%
\label{fig:Ryy}%
\end{figure}

One can define a metal and an insulator for $d\rho^{yy}/dt>0$ and $d\rho
^{yy}/dt<0$, respectively, where\ the resistivity matrix $\left\{  \rho
^{AB}\right\}  $ is the inverse of the conductivity matrix $\left\{
\sigma^{AB}\right\}  $. The metal-insulator transition in Born-Infeld
electrodynamics has been discussed in \cite{In-Cremonini:2017qwq}. So we here
focus on Logarithmic electrodynamics. In FIG. $\ref{fig:Ryy}$, we plot
$\rho^{yy}$ versus $T/\sqrt{\rho}$ for various values of $F_{yz}/\sqrt{\rho}$.
The temperature dependence of $\rho^{yy}$ is similar to the case with a larger
value of the momentum dissipation parameter in \cite{In-Cremonini:2017qwq}.
One has an insulator phase for $F_{yz}<\rho$ and a metal one for $F_{yz}>\rho
$. A metal-insulator transition could occur at $F_{yz}=\rho$, where $\rho
^{yy}=1$ is independent of the temperature.

\section{Discussion and Conclusion}

\label{Sec:Con}

In the first part of this paper, we have used the membrane paradigm to study
the electromagnetic membrane of black holes in NLED. In the membrane paradigm,
the stretched horizon is regarded as a boundary hypersurface with the surface
charge and current, which terminate the normal $D$ and tangential $H$ fields
in the region exterior to the horizon, and annul them in the interior region.
For Maxwell theory, it is well known that the horizon can be interpreted as an
ohmic conductor with a constant resistivity. It showed in
\cite{IN-Fischler:2015cma} that the horizon behaved as a Hall conductor with
surface Hall conductance\ in Maxwell-Chern-Simons theory. We derived the
conductivities of the surface current for a general NLED and found that the
conductivities usually depended on the normal electric and magnetic fields on
the stretched horizon. We also showed that there was Hall conductance for the
stretched horizon when $\mathcal{L}^{\left(  0,1\right)  }\left(  s,p\right)
$ was not zero on the horizon.

To study effects of NLED on charges scrambling on the stretched horizon, we
considered a simple scenario, in which a charged point particle freely falls
into the horizon in Rindler space. Our results showed that, during the free
falling, the surface charge density in NLED was the same as in Maxwell theory.
However, the effects of NLED would play a role in the surface current density.
In particular, when $\mathcal{L}^{\left(  0,1\right)  }\left(  s,p\right)  $
did not vanish on the horizon, there would be presence of vortices. In the
late time limit, NLED would not change the scrambling time. This is expected
since electric field becomes smaller and smaller in this limit, and we assume
that NLED would reduce to Maxwell-Chern-Simons theory for small fields. In
\cite{IN-Fischler:2015cma}, it was found that $\theta$-term only changed the
way the charge scramble but not the scrambling time in Maxwell-Chern-Simons
theory. If some NLED differs from Maxwell-Chern-Simons theory in IR limit, one
would expect the scrambling time might be changed in this NLED.

In the second part of this paper, we used the membrane paradigm approach of
\cite{IN-Iqbal:2008by} to calculate DC conductivities of an conserved current
in a field theory living on the boundary of some black brane. We assumed the
this conserved current was dual to a probe NLED field in bulk. We found that
the conjugate momentum of the NLED field encoded the information about the
conductivities both on the stretched horizon and in the boundary theory and,
in the zero frequency limit, did not evolve in the radial direction.
Therefore, we showed that these DC conductivities depended only on the
geometry and NLED fields at the black hole horizon, not on these of the whole
bulk geometry. Relating electromagnetic quantities at the horizon to these in
the boundary theory, we also showed that the DC conductivities usually
depended on the probe charge density and magnetic field in the boundary theory.

We conclude this paper with a few remarks. First, we showed that the DC
conductivities depended on the values of the couplings in NLED at the horizon.
However, authors of \cite{IN-Iqbal:2008by} showed, in Maxwell-Chern-Simons
theory, the Hall conductivity $\sigma^{yz}$ was determined by the value of
$\theta$ coupling at the boundary of the bulk. We think that the discrepancy
comes from that authors of \cite{IN-Iqbal:2008by} failed to realize that the
first term on the left-hand side of eqn. $\left(  53\right)  $ in
\cite{IN-Iqbal:2008by} was not $r$-independent any more in
Maxwell-Chern-Simons theory. On the stretched horizon, this term only
contributes to the diagonal components of the conductivities. However, this
term is now $r$-dependent and would contribute to the off-diagonal components
as well as the diagonal ones on the boundary of bulk. In other words, the Hall
conductivity of the boundary theory receives contributions from both terms on
the left-hand side of eqn. $\left(  53\right)  $, not just the first one.
These two contributions indeed make the Hall conductivity depend on the value
of $\theta$ coupling at the horizon. This incorrect statement of
\cite{IN-Iqbal:2008by} has also been noted in \cite{IN-Donos:2017mhp}, where
the authors found that the $\theta$ parameter could vanish on the boundary
with non-vanishing values on the horizon, hence giving rise to non-vanishing
Hall conductivity.

Second, one usually only turns on the electric field in the boundary theory to
calculate the holographic conductivities due to difficulties of solving the
differential equations. On the other hand, membrane paradigm provides a simple
way to obtain the dependence of holographic DC conductivities on the
electromagnetic quantities in the boundary theory, e.g. the charge density and
magnetic field. Our analysis was carried out in the long wavelength and low
frequency limit, which corresponds to an equilibrium and homogeneous state. In
particular, the charge density and magnetic field in the boundary theory are
kept fixed, time independent and homogeneous in this limit.

Finally, we only considered a neutral black brane, which is dual to a boundary
theory without a background charge density. As shown in
\cite{IN-Blake:2013bqa}, the low frequency behavior of the conductivities
depends crucially on whether there is a background charge density. It is very
interesting to study the behavior of DC conductivities in a boundary theory
with a non-vanishing background charge density, which is dual to a NLED
charged black hole.

\begin{acknowledgments}
We are grateful to Houwen Wu, Zheng Sun, Jerome Gauntlett, Aristomenis Donos,
Li Li, and Matteo Baggioli for useful discussions and valuable comments. This
work is supported in part by NSFC (Grant No. 11005016, 11175039 and 11375121).
\end{acknowledgments}

\end{document}